\newcommand{\ctbd}[1]{}

\newcommand{\lcs}{light curves}

\newcommand{\arcdeg}{\ensuremath{^{\circ}}}
\newcommand{\arcsec}{\ensuremath{^{\prime\prime}}}

\newcommand{\pxs}{\ensuremath{\rm \arcsec pixel^{-1}}}

\newcommand{\ccdsize}[1]{\ensuremath{\rm #1\times\rm#1}}

\newcommand{\kms}{\ensuremath{\rm km\,s^{-1}}}

\newcommand{\gcmc}{\ensuremath{\rm g\,cm^{-3}}}

\newcommand{\logg}{\ensuremath{\log{g}}}
\newcommand{\vsini}{\ensuremath{v \sin{i}}}

\newcommand{\sun}{\ensuremath{\odot}}

\newcommand{\msun}{\ensuremath{M_\sun}}

\newcommand{\mstar}{\ensuremath{M_\star}}

\newcommand{\rjup}{\ensuremath{R_{\rm J}}}
\newcommand{\mjup}{\ensuremath{M_{\rm J}}}

\newcommand{\figr}[1]{Fig.~\ref{fig:#1}}

\documentclass[graybox,natbib,nosecnum]{svmult}
\bibpunct{(}{)}{;}{a}{}{,} 

\pdfoutput=1   

\usepackage{mathptmx}       
\usepackage{helvet}         
\usepackage{courier}        
\usepackage{type1cm}        

\usepackage{makeidx}         
\usepackage{graphicx}        
\usepackage{multicol}        
\usepackage[bottom]{footmisc}
\usepackage[normalem]{ulem}	
\usepackage{hyperref}  

\usepackage{soul}   

\usepackage{comment}

\newcommand{\hbindex}[1]{\hl{#1}\index{#1}}  

\makeindex             

\begin{document}

\title*{The HATNet and HATSouth Exoplanet Surveys}
\author{G\'asp\'ar \'A.~Bakos}
\institute{Department of Astrophysical Sciences, Princeton University, 
	NJ 08544, \email{gbakos@astro.princeton.edu}}
%
%
\maketitle

\abstract{
The \hbindex{Hungarian-made Automated Telescope Network}
(\hbindex{HATNet}) has been in operation since 2003, with the 
key science goal
being the discovery and accurate characterization of
\hbindex{transiting extrasolar planets} (TEPs) around bright stars. 
Using six small, 11\,cm\ aperture, fully \hbindex{automated telescopes}
in Arizona and Hawaii, as of 2017 March, it has discovered and
accurately characterized 67 such objects.  The \hbindex{HATSouth}
network of telescopes has been in operation since 2009, using slightly
larger, 18\,cm diameter optical tubes.  It was the first global network
of telescopes using identical instrumentation.  With three premier
sites spread out in longitude (Chile, Namibia, Australia), the HATSouth
network permits round-the-clock observations of a 128 square arcdegree
swath of the sky at {\em any given time}, weather permitting.  As of
this writing, HATSouth has discovered 36 transiting exoplanets.  Many
of the altogether $\sim$100 HAT and HATSouth exoplanets were the first
of their kind.  They have been important contributors to the rapidly
developing field of exoplanets, motivating and influencing
\hbindex{observational techniques}, theoretical studies, and also
actively shaping future \hbindex{instrumentation} for the detection and
characterization of such objects.
}

\section{HATNet}

In 1999 we started the development of a small robotic telescope mount
with the goal of monitoring the sky for stellar variability.  The
design was inspired by the ASAS \citep{pojmanski:1997:asas} project, and
was encouraged by Bohdan Paczy\'nski from Princeton University.  The
prototype horseshoe telescope mount and clam-shell dome were completed
by July 2000, and first light was taken using a Nikon 180\,mm, f/2.8
lens and Meade Pictor $768\times512$ CCD through a Bessel $I$-band
filter from Ag\'ard, Hungary.  The telescope, named the Hungarian-made
Automated Telescope (HAT-1), was then installed on the roof of Konkoly
Observatory, Budapest, and operated from September 2000 through
December 2000.

We transported HAT-1 to Kitt Peak Observatory in January 2001, and
operated it remotely with an Apogee AP10 $2K\times2K$ CCD and Bessel
$I$-band filter until its decommissioning in November 2002
\citep{bakos:2002}.  This period of 1.5 years was very helpful in
learning how to remotely operate a complex instrument, and how to reach
robust autonomous operations with minimal supervision.

In 2001/2002 the science focus shifted to the detection of 
transiting extrasolar
planets, and the telescope system was re-designed to improve tracking,
to accommodate a larger lens, and to improve the photometric precision,
as necessitated by the detection of $<$1\% signals.  We also embarked
on building multiple units to increase sky and time-coverage.  We
tested the prototype of the new-generation mount (called HAT-5) at the
Harvard-Smithsonian Center for Astrophysics (CfA/SAO) in December 2002,
and the instrument began operations at the Fred Lawrence Whipple
Observatory (FLWO) in February 2003.  The network quickly expanded,
with the installation of two more units, HAT-6 and HAT-7 at FLWO in May
2003, the Wise HAT (WHAT) telescope at Wise Observatory, Israel, in
September 2003 (decommissioned in early 2010), and two more telescope
units (HAT-8 and HAT-9) at the Submillimeter Array of SAO, atop Mauna
Kea Observatory (MKO) in November 2003.  Finally, the last unit
(HAT-10) was installed at FLWO in November 2004.  

The initial setup (2003--2006) used Canon 110\,mm diameter f/1.8
lenses, Apogee AP10 $2K\times2K$ CCDs and $I$-band filters on all
telescopes, each yielding a $8.2\arcdeg\times8.2\arcdeg$ field-of-view
(FOV).  All CCDs were replaced with Apogee U16m \ccdsize{4K} CCDs in
September 2007, yielding a $10.6\arcdeg\times10.6\arcdeg$ field-of-view
with 9\pxs\ resolution and $\sim$2 pixel wide stellar PSFs.  At the
same time, the $I$-band filters were replaced with Bessel $R$-band
filters, and then, in September 2008, to Sloan $r$ filters.  Recently
(October 2013) the CCD on HAT-7 at FLWO was upgraded to a
back-illuminated, \ccdsize{2K} FLI camera.  We have also experimented
with achieving high precision photometry with inexpensive, consumer
grade digital single-lens reflex (DSLR) cameras, such as a Canon 60D. 
These cameras were piggy-backed on the HAT instruments in 2014--2017. 
We achieved the best DSLR-based precision so far \citep{zhang:2016}.

Altogether, the complete HAT
Network (\url{http://hatnet.org})
has been fully operational in an autonomous fashion since 2004,
for 13 years. It remained
homogeneous, using close to identical instrumentation for all 6
telescopes currently in operation.  Throughout these years, the design
principle remained consistently the same: the only off-the-shelf
components were the optics, CCDs and control computers.  All remaining
parts (dome, mount, electronics, software) were designed, built and
assembled by our team.  This led to a robust setup that we could
maintain for over a decade.  Running the HAT network necessitated on
average one per mission per site per year, either for routine
or preventative maintenance.

The search for the subtle signatures of transiting exoplanets motivated
our team to develop novel observational methods, algorithms, and
software solutions.  One example is the box least-squares (BLS) fitting
method by \citet{kovacs:2002:BLS}, a widely used algorithm for
detecting periodic, box-shaped transits.  We developed the
Trend-filtering Algorithm \citep[][]{kovacs:2005:TFA} and External
Parameter Decorrelation \citep[EPD; ][]{bakos:2010:hat11} to remove
systematic noise from the \lcs, greatly improving the detection
efficiency of transits.  We also developed the PSF-broadening method
\citep{bakos:2004:hatnet}, whereby the telescope pointing is
``drizzled'' during the exposure, thus greatly reducing the photometric
noise due to critical sampling of the stellar PSF by the
front-illuminated pixels that have a non-trivial intra-pixel
sensitivity.  We developed a novel method for deriving the astrometric
solutions for wide-field images \citep{pal:2006}, and the requirement
of high precision wide-field photometry on HAT led to the development
of the FITSH package \citep{pal:2012}. The challenges of analyzing the
time-series from HATNet and HATSouth also greatly motivated the
development of the open-source software suite VARTOOLS
\citep{hartman:2016}.

The HATNet observing strategy is to assign a primary field (i.e., one
of 838 discrete pointings on the sky) to each instrument.  The
instrument continuously observes the primary field at $3$\,min cadence
over the night so long as it is above $30^{\circ}$ elevation and not
too close to the Moon.  Typically two of the instruments in Arizona are
assigned the same fields as the instruments at Mauna Kea.  The total
time spent on a given field varies significantly, from a minimum of
$\sim 3$\,months, to several years in some cases, and with anywhere
from 2,000 to 40,000 observations collected for a field (the median
number of observations being 6000).  A total of 145 fields,
covering 34.6\% of the Northern sky, have been observed and reduced to
date.

Images from the HATNet instruments are transferred to Princeton
University, where they are reduced to trend-filtered (EPD and TFA)
light curves through a fully automated pipeline.  This pipeline performs basic
CCD calibrations, determines an astrometric solution for each image,
performs aperture photometry at the fixed location of sources from the
UCAC~4 catalog \citep{zacharias:2013}, performs an ensemble calibration
of the light curves, allowing for a smooth overall flux scaling of the
sources as a function of the color and image position, detrends each
light curve against a set of instrumental parameters (EPD), and applies
the Trend Filtering Algorithm in signal-detection mode (i.e., no
attempt is made to preserve the shapes of large amplitude astrophysical
variations).  Light curves for a total of 6\,million stars have been
generated so far in the magnitude range of $r\approx 9.5$ (saturation)
to $r = 14.5$.  The photometric precision reaches $\sim$3\,mmag at the
bright end at the 3-minute cadence.  The trend-filtered light curves
are searched for periodic transit signals using the BLS
\citep[][]{kovacs:2002:BLS} algorithm.  A variety of cuts are applied
to candidate signals to remove clear false positives (mostly pulsating
stars, eclipsing binary systems, blends with neighboring eclipsing
binaries, or cases where the detected ``transit'' is due to systematic
errors in the photometry).  Candidate transit signals are then visually
inspected by multiple team members, and the selections are collated
into a single list of transit candidates.  More than 2400 candidates
have been selected from HATNet observations to date.

\begin{figure}
\includegraphics[width=117mm]{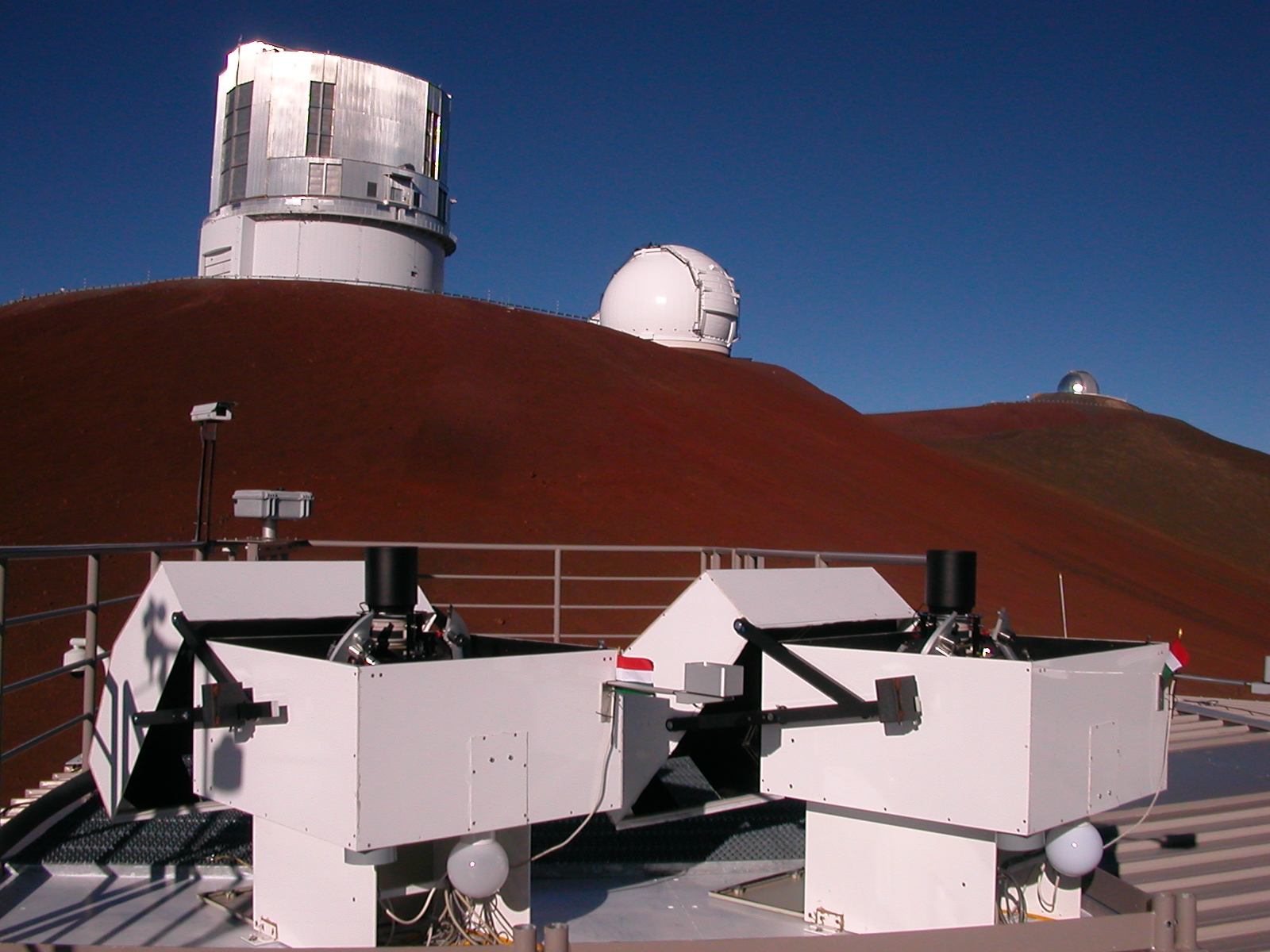}
\caption{
	Two of the HATNet telescope units at Mauna Kea, on top of the
	Submillimeter Array hangar building.  Each clamshell dome hosts a
	horseshoe mount, a 11\,cm diameter lens, and a $4K\times4K$
	front-illuminated CCD.  The systems are fully automated, and have
	been running since 2004.  The Subaru (left), Keck (center) and IRTF
	(right) telescopes are in the background.
}
\label{fig:hn}       
\end{figure}

\section{HATSouth}

\begin{figure}
\includegraphics[width=117mm]{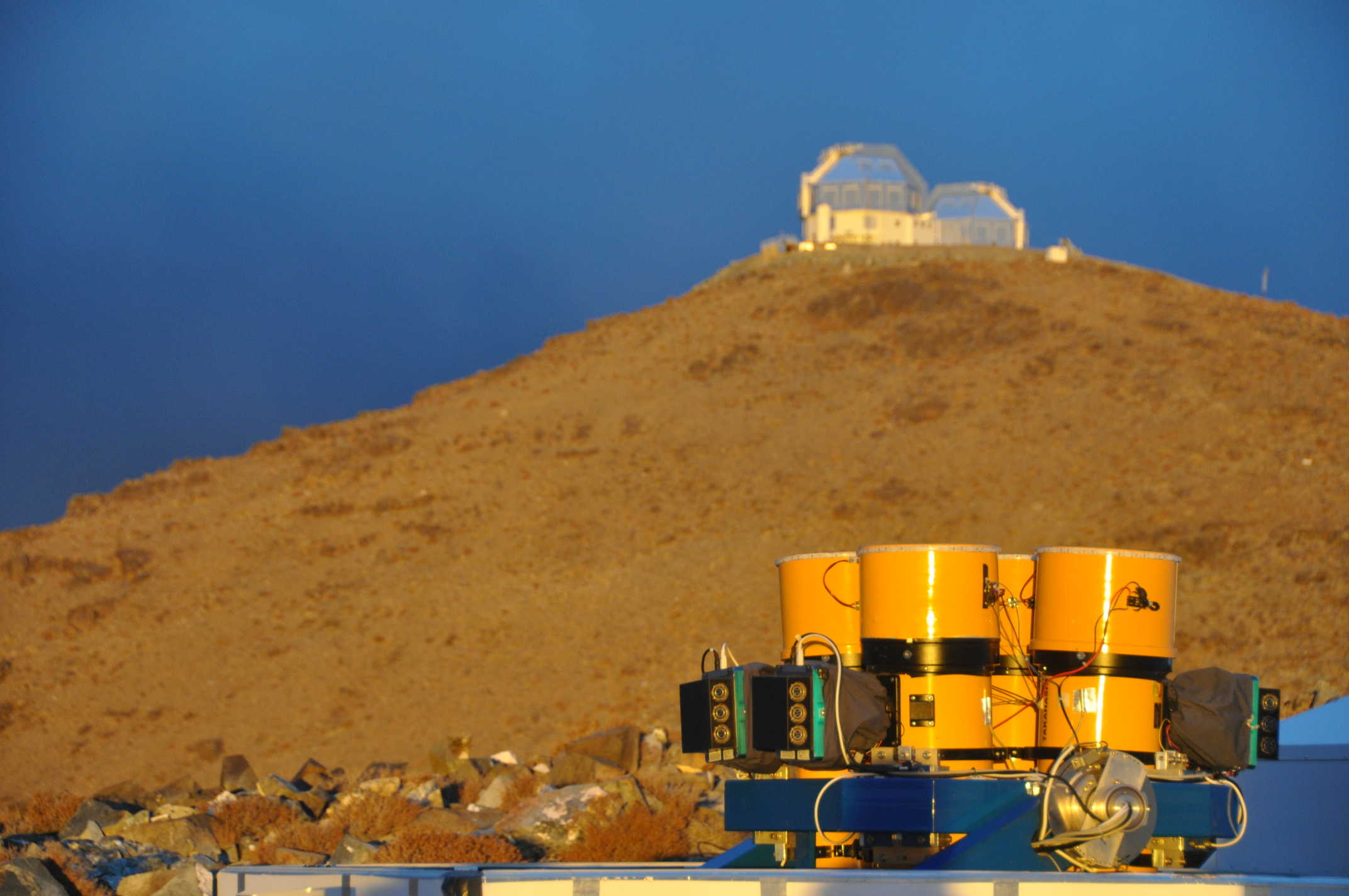}
\caption{
	One of the HATSouth telescope units at Las Campanas Observatory,
	Chile.  The diameter of the yellow optical tubes is 18\,cm.  Each
	optics is coupled with a $4K\times4K$ front-illuminated CCD.  The
	twin Magellan telescopes are in the background.
}
\label{fig:hs}       
\end{figure}

Planning of the HATSouth network (\url{http://hatsouth.org}) began in late
2007.  The prototype was fully built by the summer of 2008 at P\'ecel,
Hungary.  Altogether six ``HS$_4$'' telescope mounts were installed at
the respective sites of Las Campanas Observatory (LCO), Chile, the High
Energy Spectroscopic Survey (HESS) in Namibia, and Siding Spring
Observatory (SSO) in Australia by November 2009.  The three sites are
close to equally spread in longitude, permitting round-the-clock
observations.  Each of the HS$_4$ units (see \figr{hs}), holds four
0.18\,m diameter f/2.8 focal ratio telescope tubes on a common mount
producing an $8.2\arcdeg\times8.2\arcdeg$ FOV on the sky, imaged using
four front-side-illuminated 4K$\times$4K CCD cameras and Sloan $r'$
filters, to give a pixel scale of $3.7\pxs$.  The full HATSouth
network, with altogether 24 optical tube assemblies, was commissioned
in 2010.  In periods of good weather we have realized stretches of up
to 130 hours of non-stop observations, enabling the discovery of
many more long period transiting exoplanet candidates than can be found
by any other ground-based transit survey.  What distinguishes HATSouth
from other wide-field ground-based surveys is the fine pixel scale,
greater sensitivity to K and M dwarf host stars, and sensitivity to
long period and small radius planets.

As of March, 2017, the telescopes have opened on $\sim2300$,
$\sim2070$, and $\sim1780$ nights from LCO, HESS, and SSO,
respectively.  Based on weather statistics through 2017, the sites have
averaged 7.99\,hrs, 7.50\,hrs, and 5.48\,hrs of useful dark hours per
24\,hr time period (here dark refers to the Sun elevation below
$-12\arcdeg$). HATSouth has taken 3.35 million science
frames at 4-minute cadence, covering 17\% of the Southern sky, yielding
light curves for 10 million individual sources with $r<16$, and $\sim$4
million light curves suitable for searching for transit signals.  Data
reduction is broadly similar to that of HATNet.  To date we have
identified 1800 transiting planet candidates from the HATSouth data,
including approximately 300 with $P>10$\,days, and 60 with the size of
Neptune or smaller.

\section{From Transit Candidates to Fully Confirmed Planets}

Candidate TEPs from both HATNet and HATSouth undergo spectroscopic and
photometric follow-up observations to rule out false positives, confirm
those that are planets, and characterize their properties.  This is a
massive effort, using a plethora of facilities in a coordinated manner.

Our first follow-up step is to obtain reconnaissance moderate- to
high-resolution optical spectra for all candidates.  The spectra are
used to determine bulk properties of the star, including its radial
velocity (RV), effective temperature ($T_{\rm eff}$), surface gravity
($\logg$), metallicity and projected equatorial rotation velocity
($v\sin i$).  These are used to identify many false positives, such as
F stars transited by M dwarf stars, giants that are blended with
fainter eclipsing binary stars, or other stellar triple systems.  If
the star is {\em not} excluded by the reconnaissance spectroscopy, it
is subject to further follow-up observations.  Often the RV precision
of the reconnaissance measurements is sufficient to detect the wobble
of the star due to the orbital motion of a {\em massive} planet.  If no
RV variation is detected, then additional RV monitoring may be done
using a higher precision facility.  More than 3000 candidates have been
subject to spectroscopic follow-up, using 16 instruments altogether.
(These numbers are for HATNet and HATSouth combined). 

In parallel to and coordinated with spectroscopy, we also perform
photometric follow-up to have a higher precision light curve for
determining the system parameters, and to rule out some blended
eclipsing binary false positives through higher spatial resolution
observations.  A total of 18 facilities have been used for this effort
to gather follow-up light curves for more than 1100 candidates.  We
also regularly perform high-spatial resolution imaging for candidates
(using AO systems, speckle imaging, and lucky imaging) to search for
close stellar companions at separations less than an arcsecond.

Occasionally, additional types of follow-up are performed. This may
include spectroscopic observations taken through transit to measure the
Rossiter-McLaughlin effect \citep[e.g.,][]{zhou:2015}, or in the case
of very rapidly rotating stars, to confirm the planet via Doppler
tomography \citep[e.g.,][]{hartman:2015:hat57,zhou:2017:hatp67}.

For HAT and HATSouth, altogether, we have obtained at least some
follow-up observations for $\sim$3200 of the 4200 candidates,
confirmed $\sim140$ as substellar objects, and concluded that
$\sim2300$ are false positives or false alarms.  The remaining
candidates require more follow-up observations.

\section{Highlights}

\begin{figure}[t]
\includegraphics[width=110mm]{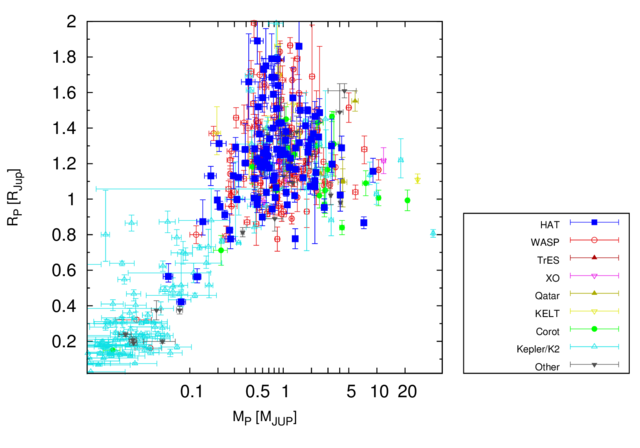}
\caption{
	The mass-radius diagram of transiting extrasolar planets, as of
	2017 March, color-coded by the projects.  HATNet and HATSouth have
	contributed exoplanets spanning a wide range of masses and radii,
	along with other physical properties.
}
\label{fig:mr}       
\end{figure}

We have been fortunate to contribute some of the exciting discoveries
to the booming field of extrasolar planets.  The physical parameters of
these exoplanets span a very wide range, as shown in e.g.~Fig.~\ref{fig:mr},
exhibiting the mass--radius diagram of those $\sim$435 exoplanets with
well measured masses and radii, and with contributions from selected
projects color-coded.  Some of the highlights from the HATNet and
HATSouth surveys include:
\begin{description}
\item[{\bf HAT-P-2b:}] the first super-massive
hot Jupiter with both mass ($9.09 \pm 0.24$\,\mjup) and radius
($1.16\pm0.073$\,\rjup) measured accurately \citep{bakos:2007:hat2}. 
\item[{\bf HAT-P-7b:}]  a short period hot Jupiter,
discovered before the launch of the {\em Kepler} mission, falling in
the field of {\em Kepler}, and later becoming one of the best studied
exoplanets \citep{pal:2008:hat7}.  It was the first planet found 
on a retrograde orbit
\citep{winn:2009:hat7,narita:2009:hat7}.
\item[\bf{HAT-P-11b:}]  the first transiting Neptune discovered by a
ground-based survey \citep{bakos:2010:hat11}.  It is also the first
Neptune with an orbital tilt measurement \citep{winn:2010:hat11} and
water vapor detection \citep{fraine:2014:hat11}.
\item[\bf{HAT-P-13b:}]  the first transiting planet in a double-planet
system with a closed orbit for the outer planet
\citep{bakos:2009:hat13}.  This system proved to be a rich dynamical
laboratory, constraining the interior structure of the inner planet
\citep{batygin:2009,buhler:2016}.
\item[\bf{HAT-P-15b:}] with $P=10.3$\,days, was the first TEP with $P >
10$\,days found by a ground-based survey \citep{kovacs:2010:hat15}. 
Along with HAT-P-17b \citep{howard:2012:hat17}, these planets were
important in realizing that the inflation of hot Jupiters is related to
the infalling flux from their host star.
\item[\bf{HAT-P-26b:}]  was the second transiting Neptune-sized planet
found by a wide field ground-based survey \citep{hartman:2011:hat26}. 
It remains one of the few Neptune-sized bodies for which transmission
spectroscopy was carried out, hinting at some properties of its
atmosphere \citep{stevenson:2016}.
\item[\bf{HAT-P-32b and -33b:}]  with $R \approx 2\rjup$ are amongst
the largest radius planets ever found \citep{hartman:2011:hat32hat33}.
\item[\bf{HAT-P-44b, -45b, and -46b:}] are
rare multi-planet transiting hot Jupiter systems 
\citep{hartman:2014:hat44hat46}. 
\item[\bf{HAT-P-47b and -48b:}] are two of the
lowest density planets ever found ($\bar{\rho} \approx 0.1\gcmc$), and are
members of the very rare class of short-period planets with masses
intermediate between Neptune and Saturn \citep{bakos:2016:hat47hat48}. 
\item[\bf{HAT-P-49b:}] orbits one of the
highest mass ($\mstar \approx 1.54\msun$) host stars found to 
date \citep{bieryla:2014:hat49}.
\item[\bf{HAT-P-54b:}]  was the first ground-based
transiting planet detection confirmed by {\em Kepler/K2}
\citep{bakos:2014:hat54}. 
\item[\bf{HAT-P-57b:}]  the fastest spinning host star ($\vsini =
100\,\kms$) with a transiting planet \citep{hartman:2015:hat57}.  This
was confirmed via Doppler tomography.
\item[\bf{HAT-P-67b:}]  is an ultra
low-density planet ($R \approx 2.1\,\rjup$, $M <
0.59$\,\mjup) around a rapidly rotating F subgiant, confirmed via
Doppler tomography \citep{zhou:2017:hatp67}.
\item[\bf{HATS-6b:}]  one of only two transiting giant planets
confirmed around an M dwarf star \citep{hartman:2015:hats6}.
\item[\bf{HATS-7b and HATS-8b:}] were, respectively, the third and
fourth transiting Super-Neptunes found by a ground-based survey
\citep{bakos:2015:hats7,bayliss:2015:hats8}.
\item[\bf{HATS-9b:}] \citep{brahm:2015:hats9}, at $t = 10.8 \pm
1.5$\,Gyr, is the oldest well characterized hot Jupiter, where the age
of the host star is known to better than 20\% precision.
\item[\bf{HATS-14b:}]  a hot Jupiter transiting a late G dwarf star
\citep{mancini:2015:hats13}, which, unlike the other such objects
known, has a high obliquity \citep{zhou:2015}.
\item[\bf{HATS-17b:}]  with $P = 16.3$\,days,
is the longest period TEP discovered by a ground-based transit survey
to date \citep{brahm:2016:hats17}.
\item[\bf{HATS-18b:}] an extremely short
period ($P = 0.84$\,day) transiting Super-Jupiter which has tidally
spun up its host star, enabling a strong constraint on the tidal
quality factor of the star \citep{penev:2016:hats18}.
\end{description}

\section{Future Prospects}
Both HATNet and HATSouth are currently running with full force. They
are providing high-precision high-cadence photometric measurements for
millions of stars, which can shed light on the long term variability of
objects, such as, for example, peculiar stars \citep{boyajian:2016}, 
or the secular variations of eclipsing binaries with period changes
\citep[e.g.][]{tylenda:2011}.  They are yielding thousands of
transiting planet candidates, which are actively followed up by an
arsenal of telescopes and instruments.  Finally, they are yielding
transiting planets at a high rate.  

Once the Transiting Extrasolar
Survey Satellite (TESS) is in operation (est.~2018), all these data
components (light curves, transiting exoplanet candidates, and
confirmed exoplanets) from HATNet and HATSouth will prove extremely
useful.  TESS will produce a high precision light curve for most bright
stars on the sky, with a typical time-span of 27 days and cadence of 30
minutes.  The spatial resolution of TESS will be twice as coarse as
HATNet, and five times worse than HATSouth.  Data from HATNet and
HATSouth will be instrumental in revealing blends; systems that mimick
transiting exoplanets, but are blends of eclipsing binaries with
brighter stars.  Archival and follow-up data from the HAT surveys will
offer further synergies with TESS: the combined data-stream will enable
the detection and characterization of transiting planets that would
otherwise be undetected in the individual survey data, and additional
astrophysics offered by the knowledge on the long term variability of
the sources.  Many of the TESS planet candidates will already have
follow-up observations through the HAT follow-up efforts.  Finally,
experience from these wide-field ground based surveys will be directly
carried over into scientific investigations with TESS.

In parallel to TESS, we anticipate that a new ground-based survey,
called HATPI, will come online.  HATPI will ultimately use an array of
63 lenses and CCDs to image the entire sky above 30 degrees (1 $\pi$
steradian) on a mosaic, every 30 seconds, whenever conditions permit,
yielding better than 3\,mmag photometric precision at 30\,s cadence for
stars at $r\approx 10$.  Construction of HATPI at Las Campanas
Observatory is under way.  The massive, high-precision data from HATPI
will complement TESS, and will also offer remarkable synergies with the Gaia
space mission.

\begin{acknowledgement}
Bakos is most grateful for the dedicated help of friends, colleagues,
past and present students and the HAT and HATSouth teams, in no
specific order:
P.~S\'ari, J.~L\'az\'ar, I.~Papp, 
J.~Hartman, K.~Penev, W.~Bhatti, M.~de Val Borro, Z.~Csubry, 
G.~Kov\'acs, R.~Noyes, D.~Sasselov, D.~Latham, G.~Torres, G.~Marcy, 
A.~Howard, D.~Fischer, J.~Johnson, T.~Henning, P.~Sackett, A.~Jord\'an, 
D.~Bayliss,
B.~Schmidt, P.~Conroy, M.~Rabus, A.~Shporer, V.~Suc, L.~Buchhave,
D.~Kipping, J.~Bento, L.~Mancini, A.~Bieryla, I.~Boisse, Z.~Tobler, 
I.~Domsa, G.~G\'alfi, B.~Cs\'ak,
A.~P\'al, B.~Sip\H{o}cz, G\'abor Kov\'acs, B.~B\'eky, G.~Zhou, Z.~M.~Zhang,
X.~Huang, M.~Soares-Furtado, J.~Hoffman, J.~Wallace, L.~Bouma, 
J.~Greco, N.~Espinoza,
R.~Brahm, P.~Sarkis, S.~Ciceri, P.~Berlind, M.~Calkins. 
Bakos wishes to acknowledge the outstanding institutional support of
Princeton University, Konkoly Observatory, the Harvard-Smithsonian
Center for Astrophysics, Pontificia Universidad Cat\'olica de Chile
(PUC), the Max Planck Institute for Astronomy (MPIA), the Australian
National University (ANU), and support from funding agencies NASA, NSF
and the Packard Foundation, and the support from the directors and
staff at Las Campanas Observatory, Siding Spring Observatory, HESS,
Fred Lawrence Whipple Observatory and the Submillimeter Array
Observatory, in running this enterprise.  HATSouth is a joint project
of Princeton, PUC, MPIA and ANU.
\end{acknowledgement}

\bibliographystyle{spbasicHBexo}  
\bibliography{bakos_hat} 

\end{document}